\documentclass[preprint,aps, showpacs, showkeys]{revtex4}

\usepackage{graphicx}

\begin{document}

\title{Statistical-mechanical description of classical test-particle
\\ dynamics in the presence of an external force field: \\modelling
noise and damping from first principles}

\author{
I. Kourakis$^{a}$
and A.P. Grecos$^{b}$
}
\affiliation{$^{a}$ Institut f\"ur Theoretische Physik IV,
Ruhr--Universit\"at Bochum, D-44780 Bochum, Germany \\
$^{b}$ Euratom - Hellenic Republic
Association, University of Thessaly, 
GR 383 34 Volos,
Greece}

\date{Submitted 10 October 2005}

\begin{abstract}
Aiming to establish a rigorous link between macroscopic random
motion (described e.g. by Langevin-type theories) and microscopic
dynamics, we have undertaken a kinetic-theoretical study of the
dynamics of a classical test-particle weakly coupled to a large
heat-bath in thermal equilibrium. Both subsystems are subject to
an external force field. From the (time-non-local) generalized
master equation a Fokker-Planck-type equation follows as a
``quasi-Markovian'' approximation. The kinetic operator thus
defined is shown to be ill-defined; in specific, it does not
preserve the positivity of the test-particle distribution function
$f(\mathbf{x}, \mathbf{v}; t)$. Adopting an alternative approach,
previously proposed for quantum open systems, is proposed to lead
to a correct kinetic operator, which yields all the expected
properties. A set of explicit expressions for the diffusion and
drift coefficients are obtained, allowing for modelling
macroscopic diffusion and dynamical friction phenomena, in terms
of an external field and intrinsic physical parameters.
\end{abstract}



\maketitle


The link between macroscopic random motion and microscopic
particle dynamics has been a long standing fundamental problem,
lying in the very heart of Non-Equilibrium Statistical Mechanics.
In a generic manner, fluctuations due to particle interactions
(\textsl{collisions}) are modelled by a Fokker-Planck-type
equation (FPE), typically related to a Langevin-type equation of
motion, which may either be derived intuitively, via physical
phenomenology or, formally, through kinetic-theoretical arguments
\cite{Van-Kampen2001a}. In the latter framework, a number of works
in Non-Equilibrium Statistical Mechanics have been devoted to the
study of the relaxation of a small subsystem (e.g. a
\textsl{test-particle}, t.p.) weakly interacting with a heat bath.
A common aim of such studies is the derivation of a {\em kinetic
equation}, describing the evolution in time of a phase-space
probability density function. This is achieved by using either
perturbation theory (typically a {\it BBGKY hierarchy} of
equations for reduced distribution functions (d.f.)
\cite{Balescu1975,Kaniadakis}) or formal theories for open systems
(e.g. projection-operator methods \cite{Van-Kampen-book}). In a
generic manner, both approaches rely on a
{\sl generalized master equation} (GME), which is obtained in 2nd
order in the (supposedly weak) inter-particle interaction; this is
a non-Markovian equation, i.e. is characterized by non-locality in
time. The kernel of the GME needs to be evaluated along particle
trajectories, so the influence of external force fields on the
microscopic laws of motion is in principle expected to modify the
form of the collision operator. 

This paper is devoted to the rigorous
derivation of a Fokker-Planck-type kinetic equation from
microscopic dynamics, taking into account the existence of an
external force field and interactions (possibly of long range)
between particles. 
This work aims addresses a number of fundamental questions arising in the description of open systems \cite{Van-Kampen2001a,Van-Kampen2001b} and, as a matter of fact,
complements previous studies relying on formal projection operator methods \cite{Grecos1985,Grecos1988,Tzanakis2}, where these issues were first considered with respect to classical systems. 

\paragraph{The model \label{secmodel}.} 

We consider a test-particle (t.p.), say $\Sigma$, surrounded by
(and weakly coupled to) a homogeneous reservoir $R$; by ${\bf X} =
({\bf x},{\bf v}) \equiv ({\bf{x_{\Sigma}}}(t),
{\bf{v_{\Sigma}}}(t)) $ and ${\bf X_{R}} \equiv \{{\bf X_{j}}\} =
\{{\bf{x_{j}}}(t), {\bf{v_{j}}}(t), \, j=1, 2, 3, ..., N\}$ we
shall denote the coordinates of the test- ($\Sigma-$) and
reservoir- ($R-$) particles, respectively. Both subsystems are
subject to an external force field, which affects the particle
dynamical trajectories $\{\mathbf{x}(t), \mathbf{v}(t)\}$. In the
absence of interactions (collisions), particle trajectories
correspond to the solution of the (single particle) dynamical
problem of motion $d^2 \mathbf{x}(t)/dt^2 = \mathbf{F_0}$.

The Hamiltonian of the system is:
\begin{equation}
H = H_{R} + H_{\Sigma} + \lambda H_{I} \, .
\end{equation}
Here $H_{R}$ ($H_{\Sigma}$) is the Hamiltonian of the
reservoir (t.p.): \(H_{R} = \sum_{j=1}^{N} H_{j} \,+
\,\sum_{j<n}\sum_{n=1}^{N}V_{jn} \). The single-particle
Hamiltonian $H_{j}$ ($j = 1, 2, ...,$ $N, \Sigma$) takes into
account the external field. $H_{I}$ stands for the interaction
(assumed to be weak: $\lambda \ll 1$) between $\Sigma$ and $R$:
\( H_{I} = \sum_{n=1}^{N}V_{\Sigma n}\),
where $V_{ij} \equiv V(|{\bf x_{i}}-{\bf x_{j}}|)$ ($i,j =1
,2,...,N,\Sigma$) is a binary-interaction potential (possibly a
long-range e.g. Coulomb-type one). The resulting equations of
motion
are:
\begin{equation}
\dot {\bf x} = {\bf v} \, \, , \qquad \dot {\bf v} = {\bf
F_{0}}({\bf x},{\bf v}) \, +\, \lambda \, {\bf F_{int}}({\bf
x},{\bf v}; {\bf X_R}; t) \label{eq:eqsmotion}
\end{equation}
\noindent The force ${\bf F_{0}}$ is due to the external field.
The {\em interaction} force \({\bf F_{int}}({\bf x},{\bf v};{\bf
X_{R}}; t) = - \frac{\partial}{\partial {\bf x}}\sum V(|{\bf
x}-{\bf x_{j}}|)\) (actually the sum of interactions between
$\Sigma-$ and N $R-$particles) represents a zero-mean Gaussian 
{\em random} process, for a reservoir in
a homogeneous equilibrium state \cite{IKthesis}.

Let us assume that the zeroth-order (\textsl{`free'}) problem of
motion -- namely (\ref{eq:eqsmotion}) for $\lambda=0$ -- yields a
known analytical solution in the form: \( 
{\bf x}^{(0)}(t) = {\bf x}+\int_{0}^{t} dt'\ {\bf v}(t') = {\bf
M}(t){\bf x}+{\bf N}(t){\bf v} \) [so ${\bf v}^{(0)}(t) = {\bf
M'}(t)\,{\bf x} +{\bf N'}(t) \,{\bf v}$;  the prime denotes differentiation with respect to time $t$], 
i.e.
\begin{equation}
{{\bf x}^{(0)}(t)\choose {\bf v}^{(0)}(t)} = \left( \matrix{{\bf
M}(t) & {\bf N}(t)\cr
                            {\bf M'}(t) & {\bf N'}(t)\cr}
\right){{\bf x}\choose {\bf v}} \, \equiv \, {\bf E}(t)\, {{\bf
x}\choose {\bf v}} \, .\label{solution}
\end{equation}
with the initial condition $ \{ {\bf x}, {\bf v} \} \equiv \{ {\bf
x}^{(0)}(0), {\bf v}^{(0)}(0) \} $, implying ${\bf E}(0)={\bf I}$ (the unit matrix). For a given dynamical problem, say in $d$ dimensions ($d=1,2,3$), the
$2d \times 2d$ matrix ${\bf E}(t)$ in (\ref{solution}) satisfies
the group property: \({\bf E}(t) {\bf E}(t') = {\bf E}(t+t')
\qquad \forall t,t' \in \Re \), implying \({\bf E}(-t) = {\bf
E}^{-1}(t)\) and a number of explicit relations among the $d
\times d$ sub-matrices $\{ {\bf M}(t), {\bf N}(t)
\}$, whose form depends on the particular aspects of the dynamical problem
 into consideration.  
This working hypothesis requires that a solution of the linearized problem of motion be known. 
Specific paradigms  obeying Eq. (\ref{solution}) include: \\
(a) linear oscillators, for which (in 1D): $M(t) = N'(t) =
\cos\omega_0 t$ and $\omega_0 N(t) = -\omega_0^{-1} M'(t) =
\sin\omega_0 t$ (here $\omega_0$ is the characteristic spring vibration eigenfrequency); \\
(b) charged particles (charge $q$, mass $m$) subject to a uniform
magnetic field $\mathbf{B} = B \hat z$ along the $z-$axis, whose
(spiral) motion obeys exactly: $\mathbf{M}= \mathbf{I}$ (hence
$\mathbf{M'}= \mathbf{0}$), ${\mathbf N'}(t)$ is the rotation
matrix:
\( {\bf N'}(t) = \left(\matrix{\cos\Omega t     & \sin\Omega t &
0\cr
                            - \sin\Omega t & \cos\Omega t    & 0\cr
                0       &       0        & 1}\right)
\) 
and ${\bf N}(t) = \int_{0}^{t} dt'\ {\bf R}(t')$, where $\Omega$ is
the {\it gyroscopic} (cyclotron) frequency $ \Omega = {q B}/{m}$
\cite{IKthesis,Kourakis}; \\
(c) the free motion (Chandrasekhar 
\cite{Chandra} or 
Rosenbluth-MacDonald-Judd \cite{RMJ}) limit, in the absence of external field(s): $\mathbf{x}(t) = \mathbf{x} + t \mathbf{v}$ 
[where $\mathbf{v}(t) = \mathbf{v} = {\rm cst.}$], viz. 
$\mathbf{M} = t^{-1} \mathbf{N} = \mathbf{I}$.


\paragraph{Statistical formulation.}

Let \( \rho = \rho (\{{\bf X},{\bf X_{R}}\};t) \) be the total
phase-space distribution function (\textsl{d.f.}), normalized to
unity: $\int d{\bf X} \, \rho = 1  $. The equation of continuity
in phase space $\it \Gamma$ reads:
\begin{equation}
\frac{\partial \rho}{\partial t}    + {\bf v_{j}} \frac {\partial
\rho} {\partial {\bf x_{j}}} + \frac {\partial}{\partial{\bf
v_{j}}} (\frac{1}{m}\,{{\bf F_{j}} \,\rho}) \,=\, 0 \, ,
\label{Liouville}
\end{equation}
where a summation over $j$ ($= 1, 2, 3, ..., N, \Sigma $) is
understood.

Defining appropriate `$s$-body'
($s = 1, 2, 3, ...$) reduced distribution functions
(\textsl{rdf}), among which the ($1-$body-) test-particle
\textsl{rdf}: \(f({\bf x},{\bf v};t) = (I, \,\rho)_{R} \equiv
\int_{\Gamma_R}\,d{\bf X_{R}} \, \rho\) (normalized to unity), and
then integrating the total ($(N+1)-$particle)
Liouville equation (\ref{Liouville}), one obtains a {\sl
BBGKY hierarchy} 
of coupled evolution equations for the \textsl{rdf}s. This standard procedure \cite{Balescu1975} has here been adapted to a test-particle
problem, namely by defining 
two types of \textsl{rdf}s, depending on whether or not the t.p. is included in the
particle cluster considered \cite{IKthesis}.
Truncating the hierarchy, one obtains
\begin{eqnarray}
(\partial _{t} -  L_{0}^{(\Sigma)})\, f({\bf X}; t) \approx  \lambda^2
\, \int d{\bf X_{1}} \, L_{I} \, g({\bf X}, {\bf X_{1}}; t) \,
\, ,
\nonumber \\
(\partial _{t} - L_{0}^{(\Sigma)} - L_{0}^{(1)} )\, g({\bf X},
{\bf X_{1}}; t)  \approx  \lambda \,
 L_{I} \, F_{1}({\bf X_{1}}) \, f({\bf X}) 
\, , \label{BBGKY-truncated}
\end{eqnarray}
where $L_{0}^{(j)}$ ($j \in \{\Sigma, 1_{R}$\}) is the ``free''
Liouvillian (in the presence of the field):
\begin{equation}
L_{0}^{(j)}\,\cdot = - {\bf v_{j}} \,\frac
{\partial\,\cdot}{\partial {\bf x_{j}}}  - \frac
{1}{m_{j}}\,\frac{\partial}{\partial {\bf v_{j}}} \,({\bf F_{0}}
\,\,\cdot\,) \label{eq:defLo}
\end{equation}
and $L_{I} \equiv L_{\Sigma1}$ is the binary interaction operator:
\begin{equation}
L_{I} = -{\bf F_{int}}(|{\bf x}-{\bf x_{1}}|) \, \biggl( \frac
{1}{m_{}} \frac {\partial} {\partial {\bf v}} - \frac {1}{m_{1}}
\frac {\partial}{\partial {\bf v_{1}}} \biggr) \, .
\label{eq:defL12}
\end{equation}
Recall that the ``tags'' $\Sigma$ (1) refer to the t.p. (one, 
any, of the reservoir particles); thus, $f = f({\bf X};
t)$, $F_{1}({\bf X_{1_{R}}})$ and $f_{2}({\bf X}, {\bf X_{1}}; t)$
denote the $\Sigma-$1-body, $R-$1-body and $(1_{R} +
\Sigma)-$2-body \textsl{rdf}s respectively, while $g = g({\bf X},
{\bf X_{1}}; t)$ is the `two-body' ($1_{R} + \Sigma$) correlation
function: \( g({\bf X}, {\bf X_{1}}; t) = f_{2}({\bf X}, {\bf
X_{1}}; t) - F_1({\bf X_{1}}) f({\bf X}; t) \). Note that the
mean-field {\sl (Vlasov)} term ($\sim \lambda^{1}$) cancels, for
reasons of symmetry, since the reservoir is assumed to be in a
homogeneous equilibrium state $F_1 = n\, \phi_{eq}({\bf v_{1}})$
($n = N/V$ is the particle density), i.e. 
$\partial F_1/\partial t = L_0^{(1)}\,F_1 = 0$.

Neglecting correlations at $t=0$, Eqs. (\ref{BBGKY-truncated})
lead to the \emph{Non-Markovian Generalized Master Equation}
(GME):
\begin{equation}
\partial _{t} f -
L_{0}\, f = \, \lambda^{2} \, n \, \int_{0}^{t} d\tau \int d{\bf
x_{1}}\,d{\bf v_{1}} \, L_{I} \, U_0(\tau) \, L_{I} \, \phi_{eq}
({\bf v_{1}}) \, f \label{eq:NMME}
\end{equation}
[here $f=f({\bf x,v};t)$], where the {\em ``free''} Liouville operator
$L_{0} \equiv L_{0}^{(\Sigma)}$ was defined in (\ref{eq:defLo}),
$L_{I}$ is the binary interaction Liouville operator $L_{\Sigma1}$
[cf. (\ref{eq:defL12})] and $U_0(\tau) =
U_0^{(\Sigma)}(\tau)\,U_0^{(1)}(\tau)$ is the evolution operator
(\textsl{propagator}) defined by the formal solution of the
``free'' (collisionless) Liouville equation [i.e.
(\ref{BBGKY-truncated}a) for $\lambda = 0$]: \( f(t) =
e^{L_0^{(j)}\,t} \,f(0) \equiv U_0^{(j)}(t) \, f(0) \) (for $j \in
\{\Sigma, 1$\}).

The procedure described
here formally amounts to defining the projection: \( {I\hskip -4.5
pt {P}}\, \cdot \, = \sigma_{R}\, \int d\Gamma_{R}\, \cdot \)\ ,
where $\sigma_{R}$ and $\Gamma_{R}$ denote the reservoir
distribution function and phase space, respectively (notice that:
\( {I\hskip -4.5 pt {P}}^2\, = {I\hskip -4.5 pt {P}} \) ), and
seeking an evolution equation for the t.p. \textsl{rdf} $f = \sigma_{R}^{-1}\, {I\hskip -4.5 pt {P}}\, \rho $. This may
be achieved by defining the complementary projection, say $Q =
{I\hskip -4.5 pt {I}} - {I\hskip -4.5 pt {P}}$ (where ${I\hskip
-4.5 pt {I}}$ is the identity operator), and then deriving a pair
of equations describing the evolution of $P \,\rho$ and $Q \,\rho$
in time \cite{Grecos1985}. This idea has been rigorously
elaborated in Refs. \cite{Grecos1988,
Tzanakis2}.

\paragraph{A `quasi-Markovian' approximation -- the $\Theta-$ operator.}

A widely used ``markovianization'' procedure consists in
substituting with the zeroth-order solution, i.e. assuming that
$f(t - \tau) \approx e^{-L_{0}\tau} \, f(t)\, \equiv \,
U_{0}(-\tau)\, f(t)$, and then evaluating the kernel
asymptotically i.e. taking the time integration limit $t$ to
infinity. A kinetic operator is thus formally defined, here
denoted as the $\Theta-$ operator.
Let us point out that the time-propagator
$U(t)$
{\em does not} commute with $\it\Gamma$-space gradients
$\partial\over{\partial{\bf v}}$, $\partial\over{\partial{\bf
x}}$; indeed,
one rigorously obtains:
\(
U_0^{(j)}(t) \frac {\partial} {\partial {\bf v_{j}}}
 U_0^{(j)}(-t) =
{\bf N_{j}^{T}}(t) \ {\partial \over
\partial {\bf x_{j}}} + {{\bf N'_{j}}}^{T}(t) \ {\partial \over \partial {\bf
v_{j}}}
\),  for $j=\Sigma,\ 1_{R}$ (a similar expression holds for 
$\partial\over{\partial{\bf x_j}}$) \cite{IKthesis}. See that the
field \emph{inevitably} enters -- via the dynamical matrices
${\mathbf N_{j}}(t)$, ... -- the collision term.

Limiting ourselves to a \emph{spatially uniform} system: $f=f({\bf v};
\,t)$ and substituting from (\ref{eq:defLo}) and (\ref{eq:defL12}) into the GME (\ref{eq:NMME}), the
`markovian' approximation described above leads to a parabolic PDE
in the form
\begin{equation} {{\partial f}
\over {\partial t}} \ + \, {1 \over m} {\bf F_{0}} {{\partial f}
\over {\partial {\bf v}}} \  = \frac{\partial } {\partial
\mathbf{v}} \, \mathbf{A} \, \frac{\partial f} {\partial
\mathbf{v}} + \frac{m}{m_1} \frac{\partial } {\partial \mathbf{v}}
(\mathbf{a} f) \, , \label{eq:QMFPE-hom0}
\end{equation}
i.e.  a
\textsl{Fokker-Planck}-type diffusion equation
\begin{equation}
{{\partial f} \over {\partial t}} \ + \, {1 \over m} {\bf F_{0}}
{{\partial f} \over {\partial {\bf v}}} \  = \ - \frac{\partial }
{\partial v_i} ({\cal F}_i \ f) + \frac{\partial^2} {\partial v_i
\, \partial v_j}(D_{ij} \ f) \,  \label{eq:QMFPE-hom}
\end{equation}
(here $i, j$ = 1, 2, 3 $\equiv v_x, v_y, v_z$). Here, ${\mathbf D} =
{\mathbf A}$ is a (positive definite) \textsl{diffusion matrix},
which is given by:
%
%
\begin{eqnarray}
{\bf {A}}
& = & \frac {n} {m^2} \int_{0}^{\infty} d\tau \int d{\bf x_{1}}
\int d{\bf v_{1}}
\, \phi_{eq}({\bf v_1}) \nonumber \\
& & 
\times {\mathbf{F_{int}}(|\mathbf{x}^{(0)}-\mathbf{x_{1}}^{(0)}|)}
\otimes
{\mathbf{F_{int}}(|\mathbf{x}^{(0)}(-\tau)-\mathbf{x_{1}}^{(0)}(-\tau)|)}\,
                      {\bf N'}^{T}(\tau)
                         \nonumber \\
& \equiv & \frac {n} {m^2} \int_{0}^{\infty} d\tau \,
               {\bf C}(t,t-\tau)\,
                               {\bf N'}^{T} (\tau)
\, . \label{coefficients-general1}
\end{eqnarray} 
Notice the explicit appearance of the Green-Kubo
coefficients (for the interaction forces $\mathbf{F_{int}}$):  diffusion coefficients are thus related to the force correlation
matrix $C_{ij} = \langle {F_{int,}}_i (t)\ {F_{int,}}_j (t-\tau)
\rangle _R$ [$=C_{ij}(\tau)$, in a stationary process], in agreement with phenomenological stochastic
theories.  
Defining the Fourier transform $\tilde V_k$ of $V(r)$, this
relation takes the form:
\begin{equation} D^{(VV)}_{rs} = \frac
{n}{m^2} (2 \pi)^d \int_{0}^{\infty} d\tau \int d{\bf v_{1}} \,
\phi_{eq}({\bf v_1}) \int d{\bf k} \, \, e^{i {\bf k\, \Delta r}}
\, k_r \, k_m \,{\tilde V}_k^2 \, {N'}_{sm}(\tau)
\label{coefficients-general2}
\end{equation}
($d=$ 3, in a 3D problem), where the exponent 
${\bf \Delta r} = {\bf r}(t) - {\bf r}(t-\tau)$ (viz. $\textbf{r}
= \textbf{x}^{(0)} - \mathbf{x_1}^{(0)}$) can be computed
by making use of (\ref{solution}); e.g. \({\bf \Delta r}(\tau) =
{\bf N}(\tau) ({\bf v}^{(0)} - {\bf v_1}^{(0)})\), as in paradigms
(b) and (c) above, where ${\bf M} = {\bf I}$ \cite{IKthesis}. 
The vector $\bf\it{F}$
in the right-hand-side (\textsl{rhs}): ${\cal F}_i = {\cal
F}_i^{(V)} \equiv (1 + \frac{m }{m_1})\, \frac{\partial
D_{ij}}{\partial v_{j}} \equiv \frac{\partial D_{ij}}{\partial
v_{j}} - \frac{m }{m_1} a_i$, represents the \textsl{dynamical
friction} force suffered by the particle, due to interactions with
its environment (the reservoir).

Summarizing so far, we have derived a Fokker-Planck
kinetic equation, involving an analytical (differential)
\emph{collision operator}; a set of exact expressions for the
\emph{diffusion coefficients} and the \emph{dynamical friction}
vector are thus obtained, to be computed for a given problem
(taking into account the external field) either analytically or
numerically; cf. e.g. in Ref. \cite{IKthesis} for case (b) above.

In the general case of an \emph{inhomogeneous} d.f. $f=f({\mathbf x},
{\mathbf v}; \,t)$, one obtains the kinetic equation
\begin{equation} {{\partial f}
\over {\partial t}} \  + {\bf v} {{\partial f} \over {\partial
{\bf x}}}  + \, {1 \over m} {\bf F_{0}} {{\partial f} \over
{\partial {\bf v}}} \  = \frac{\partial } {\partial \mathbf{v}}\,
\mathbf{D} \, \frac{\partial f} {\partial \mathbf{v}} +
\frac{\partial } {\partial \mathbf{v}}\, \mathbf{G} \,
\frac{\partial f} {\partial \mathbf{x}}  + \frac{m}{m_1}
\frac{\partial } {\partial \mathbf{v}}\, (\mathbf{a} f ) \, \equiv
\Theta_2 f \, , \label{eq:QMFPE-gen0}
\end{equation}
where the above definitions still hold; ${\bf G}$ is obtained from
\emph{rhs}(\ref{coefficients-general1}) upon ${\bf N'}^{T}
\rightarrow {\bf N}^T$. The
\emph{rhs}(\ref{eq:QMFPE-gen0}) may readily be cast in the form of a
Fokker-Planck operator [i.e. formally as in
\emph{rhs}(\ref{eq:QMFPE-hom}), yet now for 6 variables (in a 3D
problem): $i, j$ = 1, 2, ..., 6 $\equiv x, y, z, v_x, v_y, v_z$].
This 6+1 variable FPE now involves a 6D friction vector: ${\cal F}
= (0, 0, 0; {\cal F}_x^{(V)}, {\cal F}_y^{(V)}, {\cal F}_z^{(V)})$
and a $6 \times 6$ diffusion matrix, in the form:
\(
{\cal D}^{(\Theta)} = \left(
\begin{array}{cc}
{\bf 0} & \frac{1}{2} {\bf G}^T\\
\frac{1}{2} {\bf G} & {\bf A}
\end{array}
\right) \, . \label{Theta-D}
\)

Here is a crucial remark to be made. From a fundamental point of view, 
the action of a
kinetic evolution operator \emph{has to} preserve the
reality, the norm and the positivity of the d.f. $f$ (which is
defined e.g. as a normalized probability distribution function);
in formal words, a \emph{semi-group} should be defined.
Furthermore, an H-theorem should be satisfied
\cite{Balescu1975}, dictating monotonic (irreversible)
convergence towards a state of equilibrium (where an entropy
function $S \equiv - H$ attains its maximum). In fact, the diffusion matrix ${\cal D}^{(\Theta)}$, involved in the
$\Theta-$operator,  is \emph{not
positive definite}, as can be readily checked 
(see that Det${\cal D}^{(\Theta)} = 0$); therefore (\ref{eq:QMFPE-hom}) determines an
\emph{ill-defined} kinetic operator, whose action \emph{does not
preserve the positivity of the d.f.} $f$ (and neither satisfies an
H-theorem). This issue has often been overlooked in the past, 
since most studies were limited to spatially homogeneous
systems (where the problem is not
manifested, since 
$\mathbf{D}^{(VV)} = \mathbf{A}$ \emph{is}, in principle, positive
defininite).

\paragraph{Towards a Markovian FPE - the $\Phi-$operator.} 

The forementioned problem has
been known in \emph{quantum mechanics} 
\cite{Van-Kampen-book,Van-Kampen2001b,Davies1974}, despite 
its being overlooked in the \emph{classical} context 
\cite{Grecos1988,Tzanakis2}.  In his remarkable work on quantum open systems \cite{Davies1974}, 
E.B. Davies proposed a new Markovian evolution operator,  which he
formally showed to define a semigroup, and was thus psoposed as a correct description of open systems' evolution towards the equilibrium
state of a thermostat. 
``Davies' device"\cite{Van-Kampen-book} was recently
successfully applied in a (classical) plasma kinetic-theoretical context
\cite{IKthesis,Kourakis}. The main
stepstones and outcome of this method will be briefly exposed in
the following.

The alternative ($\Phi-$) operator introduced by Davies (also see in \cite{Grecos1988}), 
essentially amounts to
\begin{equation}
\Phi\,\cdot \,
= \lim_{T \rightarrow \infty} \frac{1}{2T}\int_{-T}^{T}dt'\,
\,U^{(0)}(t')\, \cdot \,U^{(0)}(-t') \label{eq:defPhi-op}
\end{equation}
(applied to the kernel of the master equation
derived above); recall that $U^{(0)}(t) \equiv \exp L^{(0)} t$ is 
the zeroth-order (collisionless) evolution operator
(propagator). 

The analytical construction of the $\Phi$ operator for a given
physical problem assumes knowledge of the exact form of the
dynamical matrices (defined above), in addition
to the reservoir equilibrium state
$\phi_{eq}(\mathbf{v_1})$ (i.e. typically a
Maxwellian). The computation may be quite tedious, yet quite
straightforward: it is, in fact, rather simplified in the case of
oscillatory test-particle (zeroth-order)
motion, as in paradigms (a) and (b) above [\emph{unlike} in
free motion (c), which is unbounded and 
characterized by a continuum spectrum, giving rise to ill-defined integrals e.g. $\lim_{T \rightarrow \infty}
\frac{1}{2T} \int_{-T}^T t^2 \,dt$]; further details, here omitted for brevity, can be found in \cite{IKthesis}.

We may briefly state the outcome of the calculations in
two typical cases, namely paradigms (a) and (b)
above, where the $\Phi$ operator was shown to
provide a well-defined FPE, which was proven to
fulfill all the necessary requirements. 
In both cases, the FP-type kinetic
equation obtained features a novel collision term, 
which accounts for
\emph{diffusion in real, position space}, in addition to
\emph{velocity} space diffusion [present, alone, in the
$\Theta-$FPE (\ref{eq:QMFPE-hom})]. Furthermore, a modified
cross-velocity-position diffusion part is obtained [replacing the ``evil''
second term in \emph{rhs}(\ref{eq:QMFPE-gen0})]; finally, the dynamical friction vector ${\cal
F}$ is modified [cf. the last terms in the two Eqs. below, which are absent in
\emph{rhs}(\ref{eq:QMFPE-gen0})]

In the case of charged particle motion in a uniform magnetic
field, the long calculation yields a kinetic equation in the form
%
%
\begin{eqnarray}
\frac {\partial f} {\partial t} \, + \, {\bf v} {{\partial f}
\over {\partial {\bf x}}} \, + \, \frac {q} {m}\, ({\bf v \times
B})\, \frac{\partial f}{\partial {\bf v}} \, = \,
  \biggl(
       \frac{\partial^{2}}{\partial v_{x}^{2}}
     + \frac{\partial^{2}}{\partial v_{y}^{2}}
  \biggr)
\biggl[
          D_{\perp} ({\bf v}) f
\biggr] \, + \, \frac{\partial^{2}}{\partial v^{2}_{z}}
                          \bigl[ D_{\parallel} ({\bf v}) f \bigr]
\qquad \qquad \qquad
\nonumber \\
+ 2\, \Omega^{-1}
    \biggl[ \frac{\partial^{2}}{\partial v_{x} \partial y} -
           \frac{\partial^{2}}{\partial v_{y} \partial x}      \biggr]
\biggl[
          D_{\perp} ({\bf v}) f
\biggr] + \, \Omega^{-2} \, \bigl[ D^{(XX)}_{\perp} ({\bf v})
\bigr] \bigl(
        \frac {\partial^{2}} {\partial x^{2}} + \frac {\partial^{2}} {\partial y^{2}}
\bigr)
f \nonumber \\
%
%
- \frac{\partial}{\partial v_{x}}
    \biggl[
            {\cal F}_{x} ({\bf v}) \, f
    \biggr]
- \frac{\partial} {\partial v_{y}}
    \biggl[
            {\cal F}_{y} ({\bf v}) \, f
    \biggr]
- \frac{\partial} {\partial v_{z}}
    \biggl[
            {\cal F}_{z} ({\bf v}) \, f
    \biggr]
+ \, \Omega^{-1} \,
     {\cal F}_{y} ({\bf v}) \, \frac{\partial} {\partial x}
  \, f
- \, \Omega^{-1} \,
     {\cal F}_{x} ({\bf v}) \, \frac{\partial} {\partial y}
  \, f \quad \quad
\label{FPE-plasma}
\end{eqnarray}
(cf. variable definitions in Sec. \ref{secmodel}); note the
cylindrical symmetry in the collision term (rhs), as imposed by
the field ($\sim \hat z$). Also note the XY-space diffusion term
in the second line.
The lengthy expressions for all coefficients (involving multiple
integrations, which have been calculated analytically to some extent, and then
computed numerically) are exposed elsewhere
\cite{IKthesis,Kourakis, IKrmf}. 
It is verified that all of the expected mathematical requirements
are fulfilled by this new plasma kinetic operator. Interestingly, 
by adopting as a working hypothesis that $D_{ij}$ and ${\cal
F}_{i}/v_i$ are constant (approximately true for low t.p.
velocities), Eq. (\ref{FPE-plasma}) models a multi-dimensional
Ornstein-Uhlenbeck random process \cite{Van-Kampen-book}, which
can be solved exactly \cite{CNSNS}.

Considering a ``1D gas'' of linear oscillators (in a parabolic
potential field $\sim \omega_0^2 x^2/2$), we have obtained:
\begin{eqnarray}
\frac {\partial f} {\partial t} \, + \, v {{\partial f} \over
{\partial x}} \, - \, \omega_{0}^{2} \,x \,\frac{\partial
f}{\partial v} \, = \, \frac{\partial^{2}}{\partial v^{2}} \biggl[
          D_{VV} (v) f
\biggr] \, +\, \frac{\partial^{2}}{\partial v \partial x} \biggl[
          D_{VX} (v) f
\biggr] \, +\, \frac{\partial^{2}}{\partial x^{2}} \biggl[
          D_{XX} (v) f
\biggr]
\, \nonumber \\
\,
%
%
- \frac{\partial}{\partial v}
    \biggl[
            {\cal F}_{V} (v) \, f
    \biggr]
%
%
- \frac{\partial}{\partial x}
    \biggl[
            {\cal F}_{X} (v) \, f
    \biggr] \, .
\label{chain2}
\end{eqnarray}
All (scalar) coefficients can be explicitly computed (relying on
the formulae derived above), yet are omitted here for brevity. The
collision term thus defined (by the \emph{rhs}) can be shown to be
well-defined as a kinetic evolution operator.

\paragraph{Conclusion - discussion.} 

Relying on first statistical mechanical principles, we have
discussed the fundamental issues involved in the derivation of a
Fokker-Planck kinetic equation (FPE), for the description of test-particle
dynamics in the presence of
an external field. Exact expressions were derived for the
diffusion and drift coefficients, which may be exactly computed
for a class of physical systems. We have thus exactly recovered the anticipated role of the Green-Kubo (force)
coefficients, relating 
diffusion coefficients to force correlations, in agreement with phenomenological stochastic
theories. The collision mechanism is nevertheless an intrinsic part of
the formalism here, and is not plainly represented by \textsl{ad
hoc} assumptions on the nature of the process. Furthermore, the external force field appears explicitly in
both correlation functions and diffusion/drift coefficients appearing in
the Fokker-Planck kinetic equation for the test-particle. The results presented here are qualitatively reminiscent of previous formal studies \cite{Grecos1985,Grecos1988,Tzanakis2}, where these issues were first discussed with respect to classical systems. 

\begin{acknowledgments} 
The authors acknowledge fruitful discussions, during the course of
this work, with L. Brenig, B. Weyssow and D. Carati (ULB,
Brussels).\\ The work of I.K. was partially supported by the Deutsche
Forschungsgemeinschaft  (Bonn, Germany) through the
{\it Sonderforschungsbereich (SFB) 591 -- Universelles Verhalten
Gleichgewichtsferner Plasmen: Heizung, Transport und
Strukturbildung}. 
\end{acknowledgments}



\begin{thebibliography}{10}

\bibitem{Van-Kampen2001a}
See the elucidating discussion in: N. G. Van-Kampen, {\it Fluct.
Noise Lett.} \textbf{1}, 3 (2001).

\bibitem{Balescu1975}
R. Balescu,\ {\it Equilibrium and Non-Equilibrium Statistical
Mechanics}, Wiley, 1975; \\ \emph{ibid},\ {\it Statistical
Dynamics}, Imperial College Press (1997).

\bibitem{Kaniadakis}
G. Kaniadakis, {\it Physics Letters A} \textbf{310}, 377 (2003).


\bibitem{Van-Kampen-book}
N. G. Van-Kampen, {\it Stochastic Processes in Physics and
Chemistry}, North-Holland, Amsterdam (1992).

%
\bibitem{Van-Kampen2001b}
N. G. Van-Kampen, {\it Fluctuation and Noise Letters}, Vol. 2
(2001) C7.

\bibitem{Grecos1985} A.P.Grecos,
in \textit{Singularities and Dynamical Systems}, S. Pnevmatikos
(ed.), North-Holland, Amsterdam (1985).

\bibitem{Grecos1988} A.P.Grecos and C.Tzanakis, Physica \textbf{A 151},
61 (1988); C. Tzanakis, A. P. Grecos, Physica 149A, 232 (1988).

\bibitem{Tzanakis2} C.Tzanakis,
\textit{On the kinetic theory of test-particles weakly-coupled to
large equilibrium systems}, PhD thesis, U.L.B. Universit\'e Libre
de Bruxelles (1987); C. Tzanakis, A. P. Grecos,
\textit{Transport Theor. Stat. Phys.} \textbf{28}, 325 (1999)
(also see in Refs. therein).

%
\bibitem{IKthesis}
I.Kourakis, {\it Kinetic Theory of Transport Processes in
Magnetized Plasmas}, PhD thesis, Universit\'e Libre de Bruxelles
(2002); online at:
\texttt{www.tp4.rub.de/$\sim$ioannis/2002DISS01.pdf}.

\bibitem{Kourakis}  I.Kourakis, {\it Plasma Phys. Control. Fusion} {\bf 41}, 587 (1999).

\bibitem{Chandra} S. Chandrasekhar, {\it Principles of Stellar
Dynamics}, University Press, Chicago (1942).

\bibitem{RMJ} M. N.
Rosenbluth, W. M. McDonald \& D. L. Judd, Phys. Rev. {\bf 107},
 1 (1957).



\bibitem{Davies1974} E.B. Davies, \textit{Comm. Math. Physics} {\bf
39},
91 (1974).
\bibitem{IKrmf} I. Kourakis, D. Carati \& B. Weyssow, {\it Proc. 2000 ICPP} (Qu\'ebec, Canada), 49 (2001); 
I. Kourakis,
{\it Revista Mexicana de F\'isica } \textbf{49} (supl. 3), 130 (2003). 


\bibitem{CNSNS} I. Kourakis and A.P. Grecos,
{\it Comm. Nonlin. Sci. Num. Sim.} \textbf{8}, 547 (2003); 
see Ch. 10 in \cite{IKthesis} for lengthy details.


\end{thebibliography}
\end{document}